\documentclass[11pt]{article}

\usepackage[english]{babel}
\usepackage[utf8]{inputenc}
\usepackage[T1]{fontenc}
\usepackage{tikz}
\usetikzlibrary{shapes.multipart}
\usepackage{lmodern}
\usepackage{url}
\usepackage{fullpage}
\usepackage{indentfirst}

\usepackage{graphics,color}
\usepackage{verbatim}
\usepackage{subfigure}
\usepackage{lineno}
\usepackage{times}

\usepackage{comment}

\usepackage{verbatim}
\usepackage{setspace, subfigure, fancyhdr}
\usepackage{amsmath,amsfonts,amssymb, bm}
\usepackage{array,dcolumn,booktabs,float}
\usepackage{url}

\def\EE{\mathbb{E}}
\def\BB{\mathbb{B}}
\def\ZZ{\mathbb{Z}}
\usepackage{graphicx}
\graphicspath{{./figuras/}}

\usepackage{float}
\usepackage{setspace}
\usepackage{natbib}
\usepackage[toc,page]{appendix}
\usepackage{wrapfig}
\usepackage[hidelinks]{hyperref}
\usepackage{bbm}

\usepackage[figuresright]{rotating}

\newcommand{\indep}{\rotatebox[origin=c]{90}{$\models$}}

\def\Y{\mathbf{Y}}

\def\X{\mathbf{X}}

\def\EE{\mathbb{E}}

\def\BB{\mathbb{B}}

\def\log{\mbox{log}}

\def\tp#1{[\![#1]\!]}

\newcommand{\tprod}[3] {
  \langle #1, #2 \rangle_{#3}}

\def\bSig\mathbf{\Sigma}

\usepackage[margin=0.8in]{geometry}

\begin{document}

\begin{singlespace}
\title{Bayesian nonparametric multiway regression for clustered binomial data}

\author{Eric F. Lock$^{1,}$\thanks{\small{\textit{Address of Correspondence}: A460 Mayo Building, MMC 303, 420 Delaware Street S.E., Minneapolis, MN 55455, USA. Tel: +1 (612) 625-0651; E-mail: \texttt{elock@umn.edu}}}  ,  Dipankar Bandyopadhyay$^{2}$\\ \\
\small{$^1$Division of Biostatistics, University of Minnesota} \\
\small{$^2$Department of Biostatistics, Virginia Commonwealth University}\\
}

\date{\ }
\maketitle
\end{singlespace}

\bigskip
\begin{singlespace}
\begin{abstract}
We introduce a Bayesian nonparametric regression model for data with multiway (tensor) structure, motivated by an application to periodontal disease (PD) data. Our outcome is the number of diseased sites measured over four different tooth types for each subject, with subject-specific covariates available as predictors.  The outcomes are not well-characterized by simple parametric models, so we use a nonparametric approach with a binomial likelihood wherein the latent probabilities are drawn from a mixture with an arbitrary number of components, analogous to a Dirichlet Process (DP). We use a flexible probit stick-breaking formulation for the component weights that allows for covariate dependence and clustering structure in the outcomes. The parameter space for this model is large and multiway: \emph{patients} $\times$ \emph{tooth types} $\times$ \emph{covariates} $\times$ \emph{components}. We reduce its effective dimensionality, and account for the multiway structure, via low-rank assumptions.  We illustrate how this can improve performance, and simplify interpretation, while still providing sufficient flexibility.  We describe a general and efficient Gibbs sampling algorithm for posterior computation.  The resulting fit to the PD data outperforms competitors, and is interpretable and well-calibrated.  An interactive visual of the predictive model is available at \url{http://ericfrazerlock.com/toothdata/ToothDisplay.html}, and the code is available at \url{https://github.com/lockEF/NonparametricMultiway}.

\vspace{12pt}
\end{abstract}
\end{singlespace}

\newpage


\section{Introduction}
Clustered, or longitudinal count data are ubiquitous in medicine and public health. Here, count responses, i.e., bio-markers for the corresponding disease, are collected in clinical studies/trials for each subject at multiple locations for the same subject (leading to clustered data), and/or at multiple occasions (time points), leading to longitudinal data. Statistical inference in this setup is typically considered under a generalized linear (or non-linear) mixed model specification, with appropriate distributional assumptions for the count responses (such as Poisson, negative binomial, etc), and the random (effect and error) terms. Our motivating data example is from a clinical study \citep{fernandes2009periodontal} of periodontal disease (PD) among Type-2 diabetic Gullah speaking African-Americans (henceforth, GAAD data), conducted at the Medical University of South Carolina. Here, the clinical attachment level (or, CAL), a clinical marker of PD, was recorded at each of the six sites of a subject's tooth producing a clustered (multivariate) setup, and we are interested in assessing the covariate-response relationships on the `number of diseased, or missing tooth sites specific to a tooth-type', such as incisors, canines, premolars, and molars. With the total count of tooth-sites $n_{ij}$ fixed for the $j$th tooth-type $(j = 1,\dots,4)$ for a subject, and the simplifying assumption that the probability of a diseased, or missing tooth site is the same within each tooth-type, this reduces to the classical binomial regression problem for clustered data, where the probability of disease/missingness can be explained by the covariates via an appropriate link-function (say, probit, logit, etc). Related estimation and inference can be easily carried out under a variety of classical and Bayesian framework using popular software, such as \texttt{SAS}, or \texttt{R}.

\begin{figure}[!h]
\centering
\includegraphics[scale=0.45]{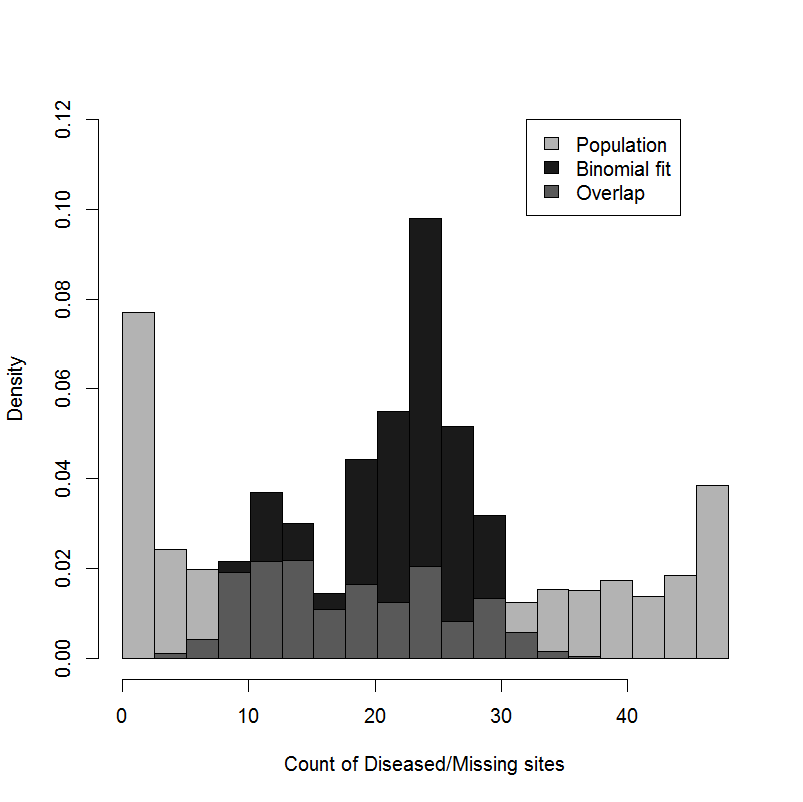}
\caption{GAAD Data: Density histogram of raw counts of diseased and missing sites (packed over tooth-types and subjects), overlayed with a Binomial fit, with the region of overlap marked.}\label{Figmotivate}
\end{figure}

However, a careful observation of the (raw) count histogram (see Figure \ref{Figmotivate}) reveals the inadequacy of a standard maximum-likelihood (ML) Binomial fit (without any covariates). The bimodal behavior of the fitted Binomial model results from $n_{ij}$ taking two values (in particular, 48 for molars, premolars and incisors; 24 for canines), and a point estimate of the binomial probability $p_{ij}$ fails to explain the rugged histogram, including the bumps at the two extremities. It is likely that an estimated $p_{ij}$ distributed as a mixture of densities in $(0,1)$ may have explained the response histogram better. However, the (automatic) determination of the \textit{number} of mixture components in a finite mixture model (FMM) is often challenging, mostly due to over- and under-fitting issues \citep{geoffrey2000finite}. Our focus is Bayesian, and most Bayesian implementation relies on popular model choice criteria, such as the Bayesian information criterion (BIC), pseudolikelihood information criterion (PLIC), etc, rendering a computationally demanding framework for searching across the cardinality spectrum of the mixture components. To alleviate this, increasingly popular Bayesian nonparametric (BNP) methods are available, where the basic idea is to fit a single model that can adapt to the observed data complexity, instead of fitting and comparing different models \citep{lai2018variational}. The most popular BNP model is the Dirichlet Process (DP) mixture model \citep{antoniak1974mixtures}, which is an infinite mixture model with a base distribution, such as the normal \citep{maceachern1998estimating}, Student's $t$ \citep{wei2012infinite}, hidden Markov models \citep{wei2011student}, etc.  The DP can be considered a special case of the class of stick-breaking priors \citep{ishwaran2001gibbs}, which define a more general family of infinite mixture models.

We consider a stick-breaking prior for the binomial probabilities $p_{ij}$, with a uniform base distribution. The DP for binomial probabilities has been considered previously \citep{gopalan1998bayesian,lock2017bayesian}; in our context, it provides a much more flexible support for the underlying distribution, and a much better fit to the observed counts (see Figure~\ref{Figmotivate}).  However, our primary interest is the relation between patient-level covariates (age, smoking status, etc.) and the proportion of diseased sites across different tooth types for each patient.  There is a large body of work on covariate-dependent modeling using Bayesian nonparametrics \citep{maceachern2000dependent,griffin2006order}.  In particular, the probit stick-breaking (PSB) formulation \citep{chung2009nonparametric,rodriguez2011nonparametric} is attractive for our context, because it admits a general framework for conditional dependence on covariates and intra-individual dependence, where the conditional distribution for the outcome is still an infinite mixture. The covariates determine the mixture weights via a probit model for each stick-breaking component. The full PSB representation involves separate coefficients for each triplet of (a) stick-breaking mixture component, (b) tooth type, and (c) covariate.  This richly parametrized model is very flexible, but is inefficient if the coefficients are treated independently. To explore a more general dependency structure among the coefficients indexed by (a)--(c) facilitated by the Bayesian proposal that automates seamless exchange of information, we cast this into a multiway regression framework \citep{lock2018tensor}  that treats the coefficients as a three-way array, and use the CANDECOMP/PARAFAC (CP) low-rank tensor factorization \citep{kolda2009tensor} for estimation. We use a similar approach to model the intra-individual correlation across components and tooth types. The use of low-rank multiway (i.e., tensor) modeling within this novel BNP context is our primary methodological contribution, which can be easily generalized to other scenarios.

In Section~\ref{model} we describe our model development, beginning with the straightforward DP model for a binomial outcome.  In Section~\ref{postcomp}, we formulate an efficient Gibbs sampling approach to posterior inference.  In Section~\ref{simulation}, we describe three simulation studies to validate our model and illustrate its motivation.  In Section~\ref{app}, we present the application to the GAAD data. Finally, in Section~\ref{discuss}, we make some concluding remarks and discuss extensions of our approach. Details on the Gibbs sampling approach are relegated to the Appendix.

\section{Statistical Model}\label{model}

\subsection{Binomial likelihood} \label{likelihood}

Let $Y_{ij}$ denote the proportion of diseased sites, and let $n_{ij}$ denote the total number of sites measured, for subject $i$ $(i=1,\hdots,I)$ in tooth type $j$ $(j=1,\hdots,J)$.  Then, \[Y_{ij} \sim \mbox{Binomial}(n_{ij}, p_{ij})\] for some latent binomial probability $p_{ij}$.  In what follows, we describe our approach to modeling the $p_{ij}$'s, beginning with a straightforward PSB model in Section~\ref{PSBP} that we extend to allow for multiway structure in Section~\ref{covmultiway}.

\subsection{Probit stick-breaking formulation} \label{PSBP}
We proceed with our BNP model development for the without- and with-covariate cases.  Without covariates, we assume the $p_{ij}$'s have a DP prior with a beta base distribution. That is, $p_{ij} \sim P$ where $P \sim DP(\mbox{Beta}(a,b),\alpha)$.  Then, each $p_{ij}$ is drawn from a theoretically infinite number of realizations $\theta_h$ from Beta$(a,b)$, with corresponding probability weights $\pi_h$,  \begin{align} p_{ij} = \sum_{h=1}^\infty \pi_h \delta_{\theta_h}, \label{pij} \end{align}  where $\delta_{\theta_h}$ is a point mass at $\theta_h$.  The distribution for the weights $\pi_h$ can be represented as a stick-breaking process \citep{sethuraman1994}, with the $\pi_h$ generated as \begin{align}\pi_h = V_h \prod_{l<h} (1-V_l),\label{eq1} \end{align} where $V_h \overset{iid}{\sim} \mbox{Beta}(1, \alpha)$. In what follows we use a uniform base distribution ($a=b=1$), and set the concentration parameter $\alpha$ to $1$ unless otherwise specified.


Now, we extend this framework to include covariates. Note that when $\alpha=1$, the distribution of the ``stick-breaks" $V_h$ (\ref{eq1}) is equivalently given by
\begin{align} V_h = \Phi(Z_{h}),\label{eq2}\end{align} where $Z_h \overset{iid}{\sim} N(0,1)$ and $\Phi$ is the standard normal CDF.  Given individual-level covariates $X: I \times D$, we extend (\ref{eq2}) to allow the stick-breaks to depend on $X$ via coefficients $\BB$, where the $h$'th column of $\BB$ gives the coefficients for component $h$ of the stick-breaking process.  That is, if  $V_{ijh}$ gives the stick-break for individual $i$, tooth type $j$ and component $h$,
\begin{align}V_{ijh} = \Phi \left(Z_{jh} +X[i,:] \BB[:,h]\right). \label{covDP} \end{align}
where, $Z_{jh}$ is an intercept term for tooth type $j$ and component $h$, $X[i,:]$ is the $i$'th row of $X$ and $\BB[:,h]$ is the $h$'th column of $\BB$. This formulation closely mimics the general approach to covariate-dependent DPs of \citet{chung2009nonparametric}, and \citet{rodriguez2011nonparametric}.

For our prior choices, we use independent $N(0,1)$ priors on the coefficients $\BB$. We also assume $Z_{jh} \overset{iid}{\sim} N(\alpha,1)$, where $\alpha \sim N(0,1)$ is a concentration parameter, such that smaller $\alpha$ favors a larger number of components with non-trivial probability under the stick-breaking process.

\subsection{Multiway extensions}\label{covmultiway}

Note that model (\ref{covDP2}) in Section~\ref{PSBP} involves three simplifying assumptions:
\begin{enumerate}
\item The effect of the individual-level covariates (age, gender, etc.), $\BB[:,h]$, are consistent across tooth types $j$ for each component $h$.
\item The coefficients $\BB$ are independent across components: $\BB[:,1] \indep \BB[:,2] \indep \BB[:,3] \cdots$.
\item The observations for each individual are independent across types $j$, given covariates $X[i,:]$.  	
\end{enumerate}

Assumption 1.\ can be relaxed with the incorporation of type-specific coefficients.  We extend model (\ref{covDP}) as follows:
\begin{align} V_{ijh} = \Phi \left(Z_{jh}+ X[i,:] \BB[:,j,h]\right), \label{covDP2}\end{align}
where $\BB[:,j,h]$ gives type-specific coefficients for each component $h$.  One ad-hoc approach to (\ref{covDP2}) is to give the entries of $\BB$ independent $N(0,1)$ priors, i.e., model each type entirely independently. Another approach is to use a standard hierarchical modeling framework across types for each component, e.g.,
\[\BB[:,j,h] \overset{iid}{\sim} N(\bar{\BB}[:,h], \Sigma) \; \text{ for } j=1,\hdots,J,\] with covariance matrix $\Sigma$. This approach assumes that covariate effects are similar across types for a given component $h$. However, the same component may have different implications by type; for example, the same atom $\theta_h$ may give a relatively high value for one type and a relatively low value for another type.  Moreover, the assumption of independent effects across the components (assumption 2.\ above) is almost certainly not satisfied.  For example, within a type covariate effects may be similar for components whose atoms are close together ($\theta_h \approx \theta_{h'})$.

With this motivation, we consider a more general model for the dependence structure of the covariate effects $\BB$. It helps to consider the full set of coefficients as the three-way array $\BB: D \times J \times H$: where $D$ is the number of individual-level covariates, $J$ is the number of types, and $H$ is the number of stick-breaking components (without truncation, $H=\infty$). Under this framework, the CP representation \citep{carroll1970analysis} restricts $\BB$ to be of low CP-rank: rank$(\BB) = R$.  That is, $\BB$ has the form of a CP factorization:
 \begin{align}
 \begin{split}\boldsymbol{\BB} =& \tp{B_1, B_2, B_3} \\
 \rightarrow  \boldsymbol{\BB}[d,j,h] =& \sum_{r=1}^R B_1[d,r] B_2[j,r] B_3[h,r]
 \label{low-rank}
 \end{split}
 \end{align}
where $B_1: D \times R$, $B_2: J \times R$, and $B_3: H \times R$ give $R$-dimensional representations for the covariates, tooth types and components, respectively.  Note that the CP factorization simply extends the traditional low-rank matrix factorization (e.g., via a singular value decomposition) to higher-order arrays.  This approach has the dual advantage of accounting for the potential multiway dependence of the coefficients, and reducing the dimensionality of the model from $J D H$ to $R (J + D + H)$.

There is a fast-growing literature on Bayesian approaches to CP factorization for multiway coefficient arrays in regression problems \citep{guhaniyogi2017bayesian,zhou2015bayesian,johndrow2017tensor}.  These approaches may be extended to our context.  In particular, we put a $N(0,1)$ prior on the marginal entries of $\BB$, as before, but have their joint pdf supported only on the space of arrays with CP-rank $\leq$ R (see Section 8 of \citet{lock2018tensor}).  This prior facilitates Gibbs sampling, and can easily be incorporated into the sampling scheme of Section~\ref{postcomp} with additional steps to sample the terms of the factorization for $\boldsymbol{\BB}$.  

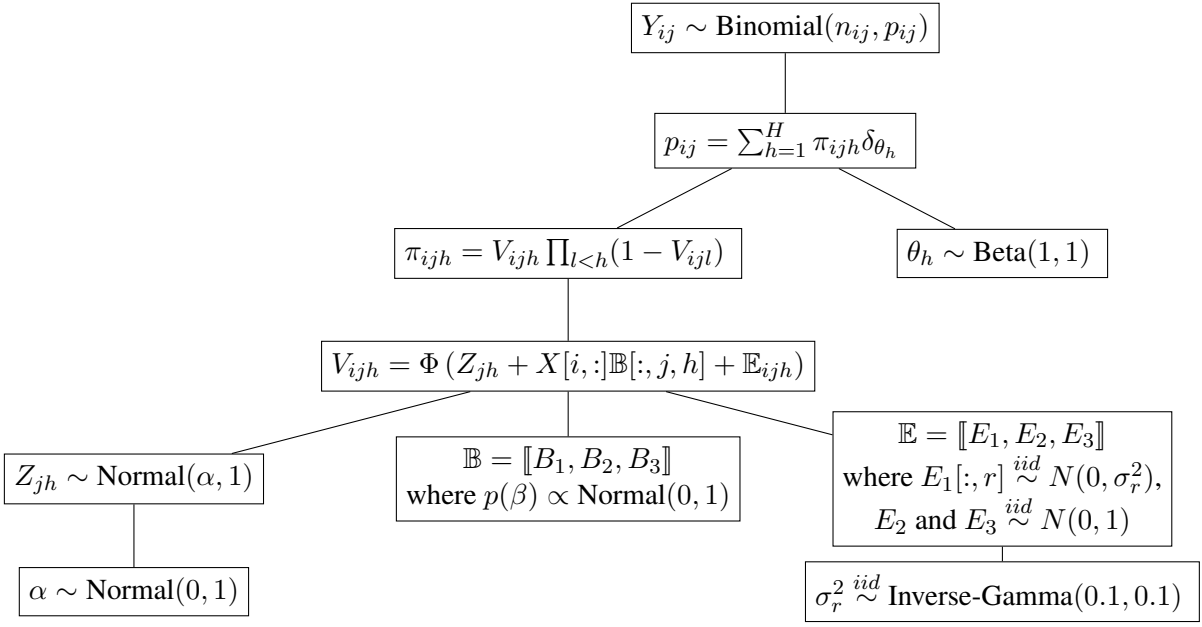
\begin{figure}
\begin{tikzpicture}[every node/.style={rectangle,draw},
 level 1/.style={sibling distance=10em},
  level 2/.style={sibling distance=15em},
  level 3/.style={sibling distance=20em},
  level 4/.style={sibling distance=15em},
  every text node part/.style={align=center}]
 \begin{centering}
\node {$Y_{ij} \sim \mbox{Binomial}(n_{ij}, p_{ij})$}
  child {node {%
  $p_{ij} = \sum_{h=1}^H \pi_{ijh} \delta_{\theta_h}$ }
  child {node {%
  $\pi_{ijh} = V_{ijh} \prod_{l<h} (1-V_{ijl})$ }
  child {node{
  $V_{ijh} = \Phi \left(Z_{jh}+ X[i,:]\BB[:,j,h]+\EE_{ijh}\right)$}
  child{ node{
  $Z_{jh} \sim \mbox{Normal}(\alpha,1)$}
  child{ node{
  $\alpha \sim \mbox{Normal}(0,1)$}}}
  child{ node{
  $\BB = \tp{B_1, B_2, B_3}$ \\
  where $p(\beta) \propto \text{Normal}(0,1)$}}
  child{ node{
  $\EE = \tp{E_1, E_2, E_3}$ \\ where $E_1[:,r] \overset{iid}{\sim} N(0, \sigma_r^2)$, \\
   $E_2$ and $E_3$ $\overset{iid}{\sim} N(0, 1)$  }
   child{ node{
  $\sigma^2_r \overset{iid}{\sim} \mbox{Inverse-Gamma}(0.1,0.1)$ } }}}}
  child {node {%
  $\theta_h \sim \mbox{Beta}(1,1)$
  }}};
  \end{centering}
\end{tikzpicture}
\caption{Hierarchical diagram of the low-rank PSB model.}
\label{diagram}
\end{figure}

Assumption 3.\ above, on the conditional independence of outcomes across tooth types, can be relaxed by incorporating correlated individual-level errors $\EE_{ijh}$:
    \begin{align} V_{ijh} = \Phi \left(Z_{jh}+ X[i,:] \BB[:,j,h]+\EE_{ijh}\right).
    \label{indivmult}
    \end{align}
These error terms can be represented as a three-way array of dimension $\EE: I \times J \times H$.  A straightforward ad-hoc approach is to model the $\EE_{ijh}$'s separately for each component $h$, with a $J \times J$ covariance matrix $\Sigma_h$ for correlation between tooth types:
\[\EE[i,:,h] \sim N(\mathbf{0}, \Sigma_h).\]
However, this model is limiting, as it does not account for the likely correlations across components $h=1,2,\hdots$, and assumes that the correlation between tooth types is consistent across components.  Instead, as for $\BB$, the low-rank CP factorization yields a flexible and parsimonious model for multiway dependence in $\EE$.   That is,
 \begin{align*}
 \EE =& \tp{E_1, E_2, E_3}
 \end{align*}
where $E_1: I \times R_e$, $E_2: J \times R_e$, and $E_3: H \times R_e$. Thus, $R_e$ latent random effects for each subject are weighted across the tooth types via $E_2$, and across the components via $E_3$.  To facilitate sampling from the posterior predictive distribution for a new subject, we consider a separable prior on each of $E_1$, $E_2$ and $E_3$, where the entries of $E_2$ and $E_3$ are independent $N(0,1)$, and the entries of \ the $r$'th column of $E_1$ (corresponding to the $r$'th latent effect) are independent $N(0,\sigma^2_r)$.  Because the scale of $E_1$, $E_2$ and $E_3$ are not independently identifiable, $\sigma^2_r$ controls the overall scale for the $r$'th rank-1 effect, allowing for appropriate shrinkage of the individual effects. We use diffuse inverse-gamma priors IG$(0.1,0.1)$ for each $\sigma^2_r$. A diagram of the full hierarchical model is shown in Figure~\ref{diagram}.

\section{Posterior Computation and Inference}
\label{postcomp}

\subsection{Augmented Gibbs sampling algorithm}
\label{gibbsMain}
The model described in Section~\ref{model} and illustrated in Figure~\ref{diagram} allows for a straightforward and efficient augmented Gibbs sampling scheme for posterior computation.  Let $C_{ij}$ give the index of the component $h$ that generated $Y_{ij}$: $C_{ij} \in \{1,\hdots,H\}$, $p_{ij} = \theta_{C_{ij}}$.  Extending the augmentation approach in \citet{albert1993bayesian}, we use latent Gaussian random variables to represent the probit link:
\[Z^*_{ijh} \sim \mbox{Normal} \left(Z_{jh}+ X[i,:]\BB[:,j,h]+\EE_{ijh}, 1\right).\]
where $C_{ij}  = h$ if and only if $Z^*_{ijh}>0$ and $Z^*_{ijh}<0$ for all $k<h$.
Gibbs sampling proceeds by sampling from the conditional distributions for
\[\{C_{ij}: i,j\}, \; \{Z^*_{ijh}: i,j,h\}, \; \{\theta_{h}: h\}, \; \{Z_{jh}: j,h\},\; B_1,\; B_2,\; B_3, \; E_1, \; E_2, \; E_3, \; \alpha,  \text{ and } \sigma^2. \]
Full details of the sampling algorithm are given in the Appendix~\ref{postcom1}.

\subsection{Posterior predictive simulation}
\label{postpred}
Given a new subject with covariates $\tilde{x}: 1 \times p$, we sample from their joint posterior predictive distribution via the generative model at each Gibbs sampling iteration.  That is, if $(\cdot)^{(t)}$ denotes the $t$'th draw for a given parameter in the Gibbs sampling chain, we draw from the posterior predictive distribution for the subject outcomes over the $J$ types $(\tilde{Y}_1, \hdots, \tilde{Y}_J)$ as follows:
\begin{enumerate}
\item Generate subject latent effects $\tilde{E}_1^{(t)} = [\tilde{E}_{11}^{(t)} \, \tilde{E}_{12}^{(t)} \, \hdots , \tilde{E}_{1R_e}^{(t)}]$ where $\tilde{E}_{1r}^{(t)} \overset{\text{indep}}{\sim} N(0,\sigma^2_r)$.
\item Determine component weights $\tilde{\pi}_{jh}^{(t)}$ for $j=1,\hdots,J$, $h=1,\hdots,H$ via
\begin{align*}
	\tilde{\EE}^{(t)} &= \tp{\tilde{E}_1^{(t)}, E_2^{(t)}, E_3^{(t)}} \\
	\tilde{V}_{jh}^{(t)} &= \Phi \left(Z_{jh}^{(t)}+\tilde{x} \BB^{(t)}[:,j,h]+\tilde{\EE}_{jh}^{(t)} \right) \\
	\tilde{\pi}_{jh}^{(t)} &=\tilde{V}_{jh}^{(t)} \prod_{l<h} (1-\tilde{V}_{jl}^{(t)}).
\end{align*}
\item Generate $\tilde{p}_j^{(t)}$ from $\sum_{h=1}^H \tilde{\pi}_{jh}^{(t)} \delta_{\theta_h^{(t)}}$ for $j=1,\hdots,J$
\item Generate $\tilde{Y}_j^{(t)}$ from $\mbox{Binomial}(\tilde{n}_{j}, \tilde{p}_{j}^{(t)})$ for $j=1,\hdots,J$.
\end{enumerate}

\subsection{Rank selection and evaluation}
\label{eval}
To select the ranks of the model, $R$ and $R_e$,  we consider the posterior predictive density of the subjects under a cross-validation scheme.  That is, we consider $p \left(Y_{i,:} \mid Y_{[i],:}\right)$,
where $Y_{i,:} = (Y_{i1}, \ldots, Y_{iJ})$ and $Y_{[i],:}$ is the data for all subjects but subject $i$. We generate samples $\tilde{p}_j^{(t)}$ from the posterior predictive distribution for $Y_{i,:}$ given $Y_{[i],:}$ as in \ref{postpred}, and then approximate the predictive likelihood as follows:
\[p \left(Y_{i,:} \mid Y_{[i],:}\right) \approx \frac{1}{T} \sum_{t=1}^T \prod_{j=1}^J {n_{ij} \choose Y_{ij}} (\tilde{p}_j^{(t)})^{Y_{ij}}  (1-\tilde{p}_j^{(t)})^{n_{ij}-Y_{ij}}. \]
To evaluate performance and select the ranks, we favor a larger log posterior predictive likelihood (LPPL):
\begin{align} \mbox{LPPL}= \sum_{i=1}^n \log \, p \left(Y_{i,:} \mid Y_{[i],:} \right). \label{LPPL}\end{align}

\section{Simulations}
\label{simulation}

Here, we describe three simulation studies, that are intended to illustrate the low-rank PSB regression model and demonstrate its flexibility.  For all studies, the generated data matched the observed GAAD data in terms of number of patient ($N=290$), number of tooth types ($M=4$), and number of patient-level covariates ($p = 6$). The first three columns of the covariate matrix $X: I \times D$ are generated independently from a $N(0,1)$ distribution, while the next three columns are generated as categorical indicators independently from a Bernoulli$(0.5)$ distribution;  following which, each column is scaled to have mean $0$ and variance $1$.  The outcomes are generated by a binomial distribution $Y_{ij} \sim \mbox{Binomial}(n_{ij}, p_{ij})$, where $n_{ij}$ is the number of sites observed for subject $i$, and tooth type $j$ in the GAAD data, and  $p_{ij}$ is generated from an assumed model. In Section~\ref{logistic}, $p_{ij}$ are given by a simple logistic model, in Section~\ref{stick-break} by a low-rank PSB regression model, while in Section~\ref{stick-breakfull} by a PSB regression model where the coefficients are not low-rank.

\subsection{Logistic simulation} \label{logistic}

We generate latent probabilities $p_{ij}$ according to the simple logistic regression model
\begin{align} \log \left(\frac{p_{ij}}{1-p_{ij}}\right) = X_{i} \beta+\epsilon_{i}, \label{logitZsim} \end{align}
where $\beta = (0,1,-1,0,1,-1)$ and $\epsilon_i \overset{iid}{\sim} N(0,1)$.  Note that the covariate effects are consistent across tooth types, and the latent error term $\epsilon_i$ is at the subject level. Thus, the underlying probabilities for a subject are the same across types, $p_{i1}=p_{i2}=p_{i3}=p_{i4}$, though the number of affected sites $Y_{ij}$ may differ.

Table~\ref{tab2} shows the LPPL (\ref{LPPL}) under various PSB models and logistic models for $p_{ij}$.  Our logistic model for inference matches the likelihood in~\eqref{logitZsim}, with diffuse independent $N(0,100)$ priors for the coefficients $\beta$, and $\epsilon_{i} \overset{iid}{\sim} N(0,\sigma^2)$ with $\sigma^2 \sim \mbox{IG}(0.1,0,1)$. We also consider analogous logistic models with independent errors ($\epsilon_{ij}$) or covariates ($\beta_j$) across tooth type: 
\[\log \left(\frac{p_{ij}}{1-p_{ij}}\right) = X_{i} \beta_j+\epsilon_{ij}.\]
  As expected, simple logistic regression models generally tend to outperform  nonparametric approaches; in particular, the logistic model with shared coefficients and errors across types matches the generative model (\ref{logitZsim}) precisely, and performs the best.  However, low-rank probit stick-breaking models are competitive. The best performing stick-breaking model has a rank-$1$ coefficient array $\BB$ and a rank-$1$ error array $\EE$.

  \begin{table}[h!]
\centering
\begin{tabular}{|c| c| c| c|}

\hline
\textbf{Stick-breaking} & No $\EE$ & Rank($\EE$) = 1 & Rank($\EE$) = 2  \\
\hline
No $\BB$ & -3958 & -3524 & -3511 \\
Rank($\BB$)=1 & -3477 & \textbf{-3135} & -3176 \\
Rank($\BB$)=2 & -3496  & -3156 & -3201 \\
Full $\BB$ &-3625 & -3324 & -3393 \\
\hline
 \textbf{Logistic} & Separate & Shared $\beta$ & Shared $\beta$, $\epsilon$\\
 & -3674 & -3628 & \textbf{-2945} \\
\hline
\end{tabular}
\caption{Mean LPPL for logistic simulation under different modeling approaches.  For the logistic models, \emph{separate} corresponds to independent coefficients and errors for each tooth type, \emph{shared} $\beta$ corresponds to shared coefficients for each tooth type, and shared $\beta, \epsilon$ corresponds to shared errors and coefficients for each tooth type.}
\label{tab2}
\end{table}

\begin{figure}[!h]
\centering
\includegraphics[scale=0.65]{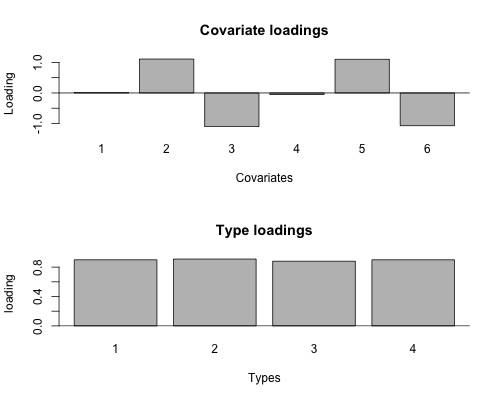}
\caption{Loadings for the covariates and tooth types for the coefficient array ($\BB$) in a rank-1 model.}
\label{rankFig}
\end{figure}

Figure~\ref{rankFig} displays elements of the fit for the best performing PSB model when rank$(\BB)=1$ and rank$(\EE)=1$, illustrating the motivation for a low-rank approach. The loadings for each covariate, $B_1$ in (\ref{low-rank}), are closely proportional to their values in the logistic model $\beta = (0,1,-1,0,1,-0)$.  The loadings for each tooth type are roughly equivalent, which makes sense, given that the covariate effects are equal across tooth types. Under the true generative model $X\beta$ suffices to describe the conditional distribution of $Y$ given $X$, \[P(Y_{ij} \mid X) = P(Y_{ij} \mid X_i \beta) \, \, \forall  \, i,. \]
Thus, the better performance of a rank-$1$ restriction on $\BB$ is explained by the appropriate reduction in dimensionality over having  a separate set of coefficients for each component of the stick-breaking process and each tooth type.  A rank-$1$ model on the error terms is also appropriate, because the unidimensional shared error term $\epsilon_i$ suffices to describe the dependence of $Y_i$ across tooth types.

\subsection{Low-rank stick breaking}
\label{stick-break}

Here, we generate data under a rank-$1$ stick breaking model.  That is, we generate data according to the stick breaking model defined by Equations~(\ref{pij}) and (\ref{eq1}), where the $\theta_h$ are generated from a Beta$(1,1)$ distribution.  The stick breaks $V_h$ are generated as in the probit formulation (\ref{covDP2}), and coefficient array $\BB$ has a rank-$1$ multiway structure (\ref{low-rank}) where the entries of the loading vectors $B_1,B_2,B_3$ are generated independently from a $N(0,1)$ distribution. Thus, under the generative model, a rank$-1$ representation for $\BB$ is correct, and the tooth types are conditionally independent given the covariates $X$.

Table~\ref{tab3} shows the LPPL (\ref{LPPL}) under different models for $p_{ij}$.  The model with Rank($\BB$)=1 and no $\EE$ closely matches the true generative model, and performs the best.  Note that the logistic models perform much worse in this context because the underlying relationship between $p_{ij}$ and $X$ does not have a logistic form, and is in general not well described by any   monotone link.  Moreover, even though the generative model is rank-$1$,  the effect of $X$ on $p_{ij}$ may differ by tooth type via different type loadings $B_2$. Hence, the logistic models with shared coefficients across tooth types perform even worse.

\begin{table}[h!]
\centering
\begin{tabular}{|c| c| c| c|}

\hline
\textbf{Stick-breaking} & No $\EE$ & Rank($\EE$) = 1 & Rank($\EE$) = 2  \\
\hline
No $\BB$ & -3965 & -3981 & -3979 \\
Rank($\BB$)=1 & -\textbf{3277} & -3301 & -3325 \\
Rank($\BB$)=2 & -3303 & -3390 & -3376 \\
Full $\BB$ &-3456 & -3477 & -3465 \\
\hline
 \textbf{Logistic} & Separate & Shared $\beta$ & Shared $\beta$, $\epsilon$\\
 & -4016 & -4214 & -4405 \\
\hline
\end{tabular}
\caption{Mean LPPL for low-rank probit simulation under different modeling approaches.  For the logistic models, \emph{separate} corresponds to independent coefficients and errors for each tooth type, \emph{shared} $\beta$ corresponds to shared coefficients for each tooth type, and shared $\beta, \epsilon$ corresponds to shared errors and coefficients for each tooth type.}
\label{tab3}
\end{table}

\subsection{Full-rank stick breaking}
\label{stick-breakfull}

Here, we generate data under a PSB model, wherein, the coefficients have no shared structure across types or components.  That is, we generate data as in Section~\ref{stick-break}, except the coefficient array that describes the stick-breaking process, $\BB$, is composed entirely of independent $N(0,1)$ entries. Thus, the underlying model has no low-rank structure, and besides the shared atoms $\theta_h$, there is no dependence among tooth types within a subject.

Table~\ref{tab3} shows the LPPL (\ref{LPPL}) under different models for $p_{ij}$.  The model with full $\BB$ and no $\EE$ closely matches the true generative model, and performs the best.  However, a rank-$2$ model has comparable performance, and this demonstrates how a model with restricted rank can reasonably approximate a more complex nonparametric model.

\begin{table}[h!]
\centering
\begin{tabular}{|c| c| c| c|}
\hline
\textbf{Stick-breaking} & no $\EE$ & Rank($Z$) = 1 & Rank($Z$) = 2  \\
\hline
No $\BB$ & -3961 & -3975 & -3981 \\
Rank($\BB$)=1 & -3401 & -3389 & -3395 \\
Rank($\BB$)=2 & -3123 & -3147 & -3167\\
Full $\BB$ &\textbf{-3105} & -3138 & -3145 \\
\hline
 \textbf{Logistic} & Separate & Shared $\beta$ & Shared $\beta$, $\epsilon$\\
 & -4005 & -4341 & -4583 \\
\hline
\end{tabular}
\caption{Mean LPPL for full probit simulation under different modeling approaches.  For the logistic models, \emph{separate} corresponds to independent coefficients and errors for each tooth type, \emph{shared} $\beta$ corresponds to shared coefficients for each tooth type, and shared $\beta, \epsilon$ corresponds to shared errors and coefficients for each tooth type.}
\label{tab3}
\end{table}

\section{Application: GAAD Data}
\label{app}

\subsection{Methods comparison} \label{comparison}

We compare the approaches described in Sections~\ref{PSBP}-\ref{covmultiway}, with alternative parametric approaches, for modeling the GAAD data.  As a parametric approach, we consider the logistic regression model
\begin{align} \log \left(\frac{p_{ij}}{1-p_{ij}}\right) = X_{ij} \beta_j, \label{logit} \end{align}
with diffuse $N(0,100)$ priors on the $\beta_j$. We also consider a model with an additional error term for unobserved factors:
\begin{align} \log \left(\frac{p_{ij}}{1-p_{ij}}\right) = X_{ij} \beta_j+\epsilon_{ij}, \label{logitZ} \end{align}
where $\epsilon_{ij} \overset{iid}{\sim} N(0,\sigma^2)$ and $\sigma^2 \sim \mbox{IG}(0.1,0.1)$.
We consider the following approaches, and compute the LPPL for each as in~\eqref{LPPL}:
\begin{enumerate}
\item \emph{Logistic model}, as in Equation~(\ref{logit}). [\textbf{LPPL:-10084}]
\item \emph{Logistic model with error}, as in Equation~(\ref{logitZ}).  [\textbf{LPPL:-4176}]
\item \emph{Marginal DP}, as described in Section~\ref{PSBP}. [\textbf{LPPL: -4222}]
\item \emph{PSB model with equal type effects}, as described in Section~\ref{PSBP}. [\textbf{LPPL: -4028}]
\item \emph{PSB model with separate type effects}, as in Equation~(\ref{covDP2}). [\textbf{LPPL: -4005}]
\item \emph{Probit DP with multiway $\BB$ and individual effects $\EE$}, as in Equation~(\ref{indivmult}): See Table~\ref{tab1}
\end{enumerate}

\begin{table}[h!]
\centering
\begin{tabular}{|c| c| c| c|}
\hline
& No $\EE$ & Rank($\EE$) = 1 & Rank($\EE$) = 2 \\
\hline
Rank($\BB$)=1 & -3953 & \textbf{-3625} &  -3645\\
Rank($\BB$)=2 & -3946  & -3647 & -3644 \\
\hline
\end{tabular}
\caption{Mean LPPL for low-rank probit DP models.}
\label{tab1}
\end{table}

Note that the best performing model in terms of LPPL is the probit DP model with rank$(\BB)=1$ and rank$(\EE)=1$.  Thus, there are clear advantages behind considering a BNP approach, and also advantages to simplifying the dimensionality of the model via low-rank constraints.  We expand on the results for this model in Sections~\ref{validation} and~\ref{interpretation}.

\subsection{Validation}
\label{validation}

\begin{figure}[!h]
\centering
  \subfigure[]{
        \includegraphics[scale=0.63]{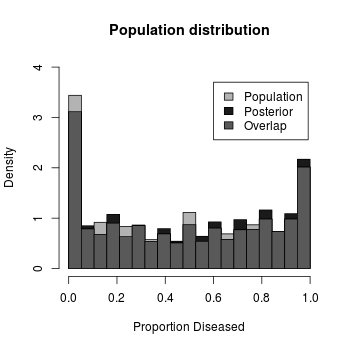} \label{fig:denscomp}}
   \subfigure[]{
        \includegraphics[scale=0.63]{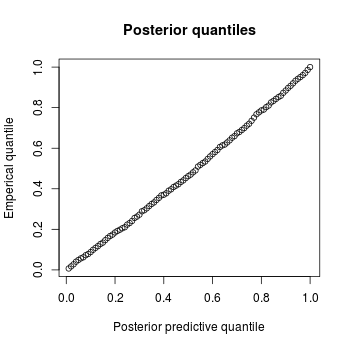} \label{fig:QQ}}
\caption{Assessment of the probit DP model with rank$(\BB)=1$ and rank$(\EE)=1$, via posterior predictive simulations under 10-fold cross-validation.  Panel (a) compares the overall distribution of posterior predictive proportions with the population distribution.  Panel (b) shows the empirical quantile plot for posterior predictive quantiles. }
\label{fig:validation}
\end{figure}

We perform a 10-fold cross-validation study of the PSB model with rank$(\BB)=1$ and rank$(\EE)=1$ to assess the calibration and coverage of the predictive model.  We randomly partition the data for $N=290$ individuals into $10$ equal subgroups of size $29$, $\{\Y_1,\X_1, \hdots, \Y_{10}, \X_{10}\}$. For each subgroup,  we draw $1000$ samples from the posterior predictive distribution of $\Y_i$, given their individual-level covariates and data for the remaining subgroups:
\[\Y_i^{(1)},\hdots,\Y_i^{(1000)} \overset{iid}{\sim} p \left(\Y_i \mid \X_i, \{\Y_{j}, \X_j\}_{j \neq i}\right). \]
Figure~\ref{fig:denscomp} shows a histogram of the proportion of diseased sites for all out-of-sample posterior predictive simulations and for the observed proportions. The marginal distribution of the simulated values closely matches that for the observed values.  To assess the calibration of posterior predictive distributions and their inferential accuracy, we consider the quantile of each observed value under its out-of-sample posterior predictive distribution $\phi_{ij} = \frac{1}{1000} \sum_{t=1}^{1000} \mathbbm{1}_{\{Y_{ij}<Y_{ij}^{(t)}\}}$, and compute the empirical CDF of the posterior predictive quantiles, as follows:
\[\hat{P}(\phi<x) = \frac{1}{290*4}\sum_{i=1}^{290} \sum_{j=1}^4 \mathbbm{1}_{\{\phi_{ij}<x\}} .\]

Figure~\ref{fig:QQ} shows a plot of the empirical CDF of the posterior predictive quantiles.  This is a straight line approximating the identity function, demonstrating that the predictive model is well-calibrated, and does not under-estimate or over-estimate uncertainty.

\subsection{Interpretation}
\label{interpretation}

\begin{figure}[!h]
\centering
  \subfigure[]{
        \includegraphics[scale=0.63]{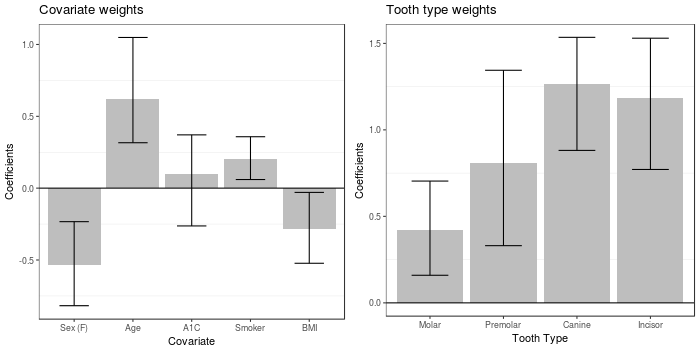}\label{fig:weights}}
   \subfigure[]{
        \includegraphics[scale=0.63]{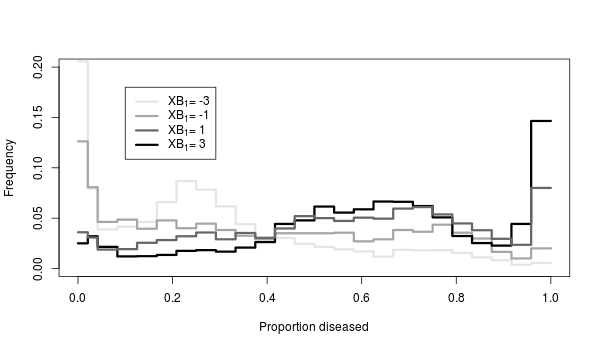} \label{fig:continuum}}
\caption{Panel (a) gives the weights for covariates and tooth types, with 95\% credible intervals, for $\BB$ when rank$(\BB)=1$ and rank$(\EE)=1$.  Panel (b) displays the change in the marginal posterior predictive distribution for different values of the covariate score $X B_1$.  }
\label{fig:rank1weights}
\end{figure}

Figure~\ref{fig:rank1weights} summarizes the posterior for the PSB model with rank$(\BB)=1$ and rank$(\EE)=1$, conditioning on the full dataset with $N=290$ individuals.  Because rank$(\BB)=1$, the covariate effects can be decomposed into a single vector of weights for each covariate, for each tooth type, and for each component of the stick-breaking process.  Figure~\ref{fig:weights} shows the weights for covariates and tooth types, with $95\%$ credible intervals.  The covariate weights show a strong positive effect of age and a weak positive effect of smoking, a strong negative effect of sex (F) and a weak negative effect of BMI, and a negligible effect of A1C.  The tooth type weights are all consistently positive, demonstrating that the effect of the covariates are similar across types, but may vary in magnitude.  The conditional distribution of diseased sites across tooth types for an individual depends entirely on their univariate covariate score $XB_1$. Figure~\ref{fig:continuum} illustrates the marginal distribution across tooth types for certain values on this continuum;  for a higher covariate score, the resulting distribution favors a higher proportion of diseased sites.  Thus, status is improved for females, non-smokers, younger individuals, and those with a higher BMI, and these effects are generally consistent across tooth types.  The findings for sex, smoking and age are consistent with the literature on periodontal disease \citep{chambrone2010predictors}.  The finding for BMI is somewhat counter-intuitive, and results are mixed in the literature \citep{kongstad2009relationship,chaffee2010association};  an ad-hoc correlation analysis of these data similarly yields a negative correlation between BMI and proportion of diseased sites for all tooth types, with p-value$<0.01$ for canines and incisors. 

Note that the use of a low-rank tensor model simplifies interpretation substantially over (e.g.) a model that allows the covariate coefficients to vary freely over different components of the PSB process and different tooth types. An online interactive visualization showing the posterior predictive distribution for proportion of diseased sites for any given tooth type and combination of individual-level predictors is available at  \url{http://ericfrazerlock.com/toothdata/ToothDisplay.html}.

\section{Discussion}\label{discuss}
For our motivating application to the GAAD data, our low-rank PSB model provides sufficient flexibility without sacrificing parsimony, and inherently accommodates the multiway outcome.  While we have focused on our motivating application with a binomial likelihood, the methodological ideas can be extended to other contexts.  For example, the model easily extends to other likelihoods; the hierarchical prior for the weights $\pi$ (Figure~\ref{diagram}) and steps 3-12 of the Gibbs sampling algorithm in the Appendix~\ref{postcom1} remain the same.  Moreover, while in our context the outcomes are a $2$-way array (patients $\times$ tooth types), an analogous approach may be used when the outcomes or predictors are of higher order.  Even if the data do not take the form of a multiway array, one can still use a PSB formulation where the coefficients (components $\times$ covariates) have a low-rank matrix factorization.  Such an approach may be particularly useful for variable selection problems, in which some covariates have no effect on any of the component weights \citep{chung2009nonparametric}.  For this, one could use a sparsity-inducing prior, such as a spike-and-slab \citep{ishwaran2005spike} or horseshoe \citep{carvalho2010horseshoe} prior, directly on the covariate loadings in the low-rank factorization (e.g., $B_1$ in (\ref{low-rank})). All the above are possible avenues for future research, and will be considered elsewhere.

\section*{Acknowledgments}
The authors thank the Center for Oral Health Research at the Medical University of South Carolina for providing the motivating data. Lock and Bandyopadhyay's research were supported by grants ULI RR033183/KL2 RR0333182 and R01DE024984, respectively, from the US National Institutes of Health.

\clearpage
\addcontentsline{toc}{section}{References}

\begin{singlespace}
\bibliographystyle{rss}
\bibliography{biblio}
\end{singlespace}

\appendix
\section{Gibbs sampling algorithm}
\label{postcom1}

The Gibbs sampling algorithm for the full model given in Figure~\ref{diagram}, using the parameter augmentation described in Section~\ref{postcomp}, proceeds as follows:
\begin{enumerate}
\item Update $C_{ij}$'s, where
\[P \left(C_{ij}=h \mid \theta_h, Y_{ij}\right) \propto \pi_{ijh}  \theta_h^{Y_{ij}} (1-\theta_h)^{n_{ij}-Y_{ij}}.\]

 \item Update the atoms $\theta_h$ for $h=1,\hdots,H$. Let $\tilde{Y}_h$ and $\tilde{n}_h$ be the total number of affected sites and total number of sites, respectively, that belong to observations allocated to DP component $j$:
\[  \tilde{Y}_h = \sum_{\{i,j: \, C_{ij}=h\}} Y_{ij} \; \;, \;\; \tilde{n}_h = \sum_{\{i,j: \, C_{ij}=h\}} n_{ij} .\]
The full conditional  distribution of $\theta_h$ is then
 \[\mbox{Beta} \left(a+\tilde{Y}_h,b+\tilde{n}_h-\tilde{Y}_h\right).\]

 \item Update the $Z^*_{ijh}$'s:  \[Z^*_{ijh} \sim \begin{cases} N(Z_{jh}+ X[i,:]\BB[:,j,h]+\EE_{ijh}
, 1)\mathbbm{1}_{\mathbb{R}^-} \; \text{ if } h<C_{ij}  \\
N(Z_{jh}+ X_{i}\beta_{jh}+\EE_{ijh}
, 1)\mathbbm{1}_{\mathbb{R}^+} \; \text{ if }  h=C_{ij} \end{cases}\]
 where $N(\cdot,1)\mathbf{1}_{\mathbb{R}^-}$ and $N(\cdot,1)\mathbf{1}_{\mathbb{R}^+}$ define a normal distribution truncated below or above $0$, respectively.
\item Update the coefficient tensor  $\BB=\tp{B_1,B_2,B_3}$.  Define the array $\ZZ^*_B$ where $\ZZ^*_B[i, j, h] = Z^*_{ijh}-Z_{jh}-\mathbb{E}_{ijh}$. Then, $\ZZ^*_B = \tprod{X}{\BB}{1}$ where $\ZZ^*_B: I \times J \times H$, $X: I \times D$, and $\tprod{\cdot}{\cdot}{1}$ defines the contracted tensor product over dimension $D$. We  sample from the full conditional distributions of $B_1$, $B_2$, and $B_3$ as in Section 8 of \citet{lock2018tensor}.

\item Update the individual effect tensor  $\EE=\tp{E_1,E_2,E_3}$.  Define the array $\ZZ^*_E$ where $\ZZ^*_E[i, j, h] = Z^*_{ijh}-Z_{jh}-X[i,:]\BB[:,j,h]$. The entries of $\ZZ^*_E$ are independent and normally distributed with mean array $\EE$ and variance $1$.  The full conditional distributions for $E_1$, $E_2$ and $E_3$ can each be found by an application of well-known results for the Bayesian linear model \citep{lindley1972bayes}.  For example, consider $E_1$. Let the columns of $E_{23}: JH \times R_e$ be given by the outer product of the respective columns in $E_2$ and $E_3$:
\[E_{23}[:,r] = \mbox{vec}\left(E_2[:,r] \otimes E_3[:,r]\right),\]
where $\mbox{vec}$ is the vectorization operator.  Let $Z^{*\text{mat}}_E: I \times JH$ give the matricization of $\ZZ^*_E$, i.e.,
\[Z^{*\text{mat}}_E[i,:] = \mbox{vec}(\ZZ^*_E[i,:,:]).\]
Then, $E_1$ regresses $Z^{*\text{mat}}_E$ on $E_{23}$, giving
\[Z^{*\text{mat}}_E[i,:] \sim \mbox{Normal} \left(Z^{*\text{mat}}_E[i,:] E_{23} \left(E_{23}^T E_{23}+\Sigma_{E_1} \right)^{-1},\left(E_{23}^T E_{23}+\Sigma_{E_1} \right)^{-1} \right)
 \; \text{ for } i=1,\hdots,I,\]
 where $\Sigma_{E_1}$ is the diagonal covariance matrix with $\Sigma_{E_1}[r,r] = \sigma^2_r$.  The full conditional distributions for $E_2$ and $E_3$ are analogous to that for $E_1$, but with diagonal prior covariance $\Sigma_{E_2}[r,r] = \Sigma_{E_3}[r,r] = 1$.
\item Update the $\sigma^2_r$'s: $\sigma_r^2 \sim \mbox{Inverse-Gamma} \left(0.1+\frac{1}{2}I, 0.1+{frac}{1}{2}\sum_{i=1}^I E_1[i,r]^2\right)$

\item Update the $Z_{jh}$'s.  Define the array $\ZZ^*_Z$ where $\ZZ^*_Z[i, j, h] = Z^*_{ijh}-X[i,:]\BB[:,j,h]-\mathbb{E}_{ijh}$.
\[Z_{jh} \sim \mbox{Normal}\left(\frac{\alpha + \sum_{i=1}^I \ZZ^*_Z[i,j,h]}{I+1}, \frac{1}{I+1} \right) \]
\item Update $\alpha$: $\alpha \sim \mbox{Normal}\left(\frac{\sum_{j=1}^J \sum_{h=1}^H Z_{jh}}{JH+1}, \frac{1}{JH+1}\right)$ for $r=1,\hdots,R_e$.

\item Update the stick-breaks $V_{ijh}$ for $h=1,\hdots,H$.  For $h=1,\hdots,H-1$ is
\[V_{ijh} \overset{iid}{\sim} \Phi \left(Z_{jh} + X_{ij}\beta_h\right),\]
 with $V_{ijH}=1$.
 \item Update the weights $\pi_{ijh} = V_{ijh} \prod_{l<h} (1-V_{ijl})$ for $h=1,\hdots,H$.
 \item Update $p_{ij} = \theta_{C_{ij}}$ for each pair $(i,j)$.
 \item Determine component weights $\pi_{ijh}$ via
\begin{align*}
	V_{ijh} &= \Phi \left(Z_{jh}+ X[i,:]\BB[:,j,h]+\EE_{ijh} \right) \\
	\pi_{ijh} &=V_{ijh} \prod_{l<h} (1-V_{ijl}).
\end{align*}

\end{enumerate}

\end{document}